\documentclass[aps,twocolumn,amsmath,amssymb,prb]{revtex4}

\usepackage{graphicx}
\usepackage{bm}

\newcommand{\bsube}{\begin{subequations}}
\newcommand{\esube}{\end{subequations}}
\newcommand{\Eqs}[1]{Eqs.\,(\ref{#1})}
\newcommand{\Eq}[1]{Eq.\,(\ref{#1})}
\newcommand{\Fig}[1]{Fig.\,\ref{#1}}
\newcommand{\be}{\begin{equation}}
\newcommand{\ee}{\end{equation}}
\newcommand{\nl}{\nonumber \\}
\newcommand{\la}{\langle}
\newcommand{\ra}{\rangle}
\newcommand{\lla}{\langle \langle}
\newcommand{\rra}{\rangle \rangle}

\newcommand{\bra}[1]{\la #1|}
\newcommand{\ket}[1]{|#1\ra}

\begin{document}

\title{Non-Markovian theory for the waiting time distributions of single electron transfers}

\author{Sven Welack}
\affiliation{Department of Chemistry, Hong Kong University of
    Science and Technology, Kowloon, Hong Kong}

\author{YiJing Yan}
\affiliation{Department of Chemistry, Hong Kong University of
    Science and Technology, Kowloon, Hong Kong}

\begin{abstract}
We derive a non-Markovian theory for waiting time distributions of
consecutive single electron transfer events.
The presented microscopic Pauli rate equation
formalism couples the open electrodes to the many-body system,
allowing to take finite bias and temperature into consideration. Numerical
results reveal transient oscillations of distinct system frequencies
due to memory in the waiting time distributions. Memory
effects can be approximated by an expansion in non-Markovian
corrections. This method is employed to calculate memory landscapes displaying
preservation of memory over multiple consecutive
electron transfers.
\end{abstract}


\maketitle

\section{Introduction}

Detection of single electron transfers through
quantum systems such as quantum dots
has become experimentally
feasible.\cite{Lu03422,Fuj042343,Gus06076605,Fuj061634}
Theoretical investigations on the underlying statistics
were mostly obtained in terms of higher cumulants,
e.g.\ noise and skewness, by using generating function
techniques.\cite{Lev964845,Lev04115305,Wab05165347,Ram04115327,%
She03485,Fli05475,Uts06086803, Ped05195330,Bac011317,Kie06033312}
Expansion of the higher cumulants
in non-Markovian corrections has revealed significant memory
effects in quantum dots \cite{Eng04136602} when
strong Coulomb interaction,\cite{Bra06026805}
phonon bath \cite{Bra081745} or initial
correlations \cite{Fli08150601} are present.

Statistics based on waiting time distribution (WTD) provides additional
information on the system. Higher cumulants can be
derived from the WTD \cite{Bra08477}, but not vice versa.
Waiting times were recently utilized to analyze single electron transfers
in the Markovian regime, for example, in double quantum
dots,\cite{Wel08195315} single molecules,\cite{Koc05056801,Wel081137}
single particle transport \cite{Bra08477} and Aharonov-Bohm
interferometers.\cite{Wel0957008}
Non-Markovian treatment of WTD has shown significant
features in photon counting statistics.\cite{Zai873897}

Non-Markovian effects are induced by a small bias voltage
or a finite bandwidth of the system-electrode coupling.
While the former can be eliminated
easily, the latter scenario is given by the setup of experiment.
In order to explore both regimes, a non-Markovian Pauli rate equation
based on a microscopic description of the electrode-system coupling using a
Lorentzian spectral density is derived. It can be utilized
for a variety of system, such as single molecule and quantum dots.

A formal connection of the WTD
with the shot noise spectrum of electron
transports through quantum junctions has been established in the
Markovian regime.\cite{Bra08477} The non-Markovian shot noise
spectrum \cite{Jin0806,Eng04136602} provides a more accurate
description of the signal
and its relation to the physics of the
junction than a Markovian version,
since it reveals several distinct intrinsic system frequencies.

\begin{figure}
\includegraphics[width=7.0cm,clip]{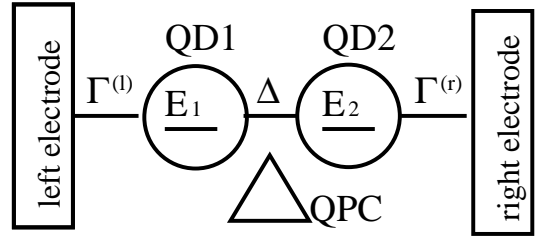}
\caption{Illustrated set-up of the DQD in series and notation
of the important parameters. The charge state of the DQD is measured
by the quantum point contact (QPC) which provides an electron transfer trajectory.
\label{fig:0}}
\end{figure}

In this paper we derive a non-Markovian theory for the WTD
of single particle transfer trajectories based on the derivation
of a non-Markovian microscopic Pauli rate equation.
It provides a general framework to study non-Markovian electron transport through many-body systems
and allows us to distinguish between non-Markovian effects due to intrinsic properties of the
system, finite electrode-system coupling band-width and small bias voltage.
The WTD is evaluated in time domain by
perturbation theory leading to non-Markovian corrections.\cite{Bra06026805}
We shall analyze the effect of memory on consecutive electron
transfers through double quantum junctions (DQD), see \Fig{fig:0},
and demonstrate the influence of many-body Coulomb coupling
on memory landscapes displaying the memory that is preserved in the system for several consecutive
electron transfers. The non-Markovian spectrum is obtained from a Laplace transformation.
The results reveal that the non-Markovian spectrum of the WTD
provides similar information content which qualifies it as an alternative method
to the non-Markovian shot noise spectrum.

The paper is organized as follows.
In section \ref{thrate}, we present the derivation of the
non-Markovian rate equation. The expressions for the
non-Markovian WTD are shown in
section \ref{secWTD}. The formalism is applied to the DQD system
and the results are given in section \ref{demo}.
We conclude with a summary and outlook.

\section{Non-Markovian rate theory of quantum transport}\label{thrate}

\subsection{Hamiltonian}

Consider a junction consisting of a DQD in
series as the system, two electron reservoirs,
and the respective system--reservoir coupling, as shown
in \Fig{fig:0}. The total Hamiltonian assumes
$H_T=H_S+H_R+H_{SR}$.
The system part describes the DQD which is modeled by
\begin{equation}\label{eq1}
H_S=\sum_{s=1}^2 E_{s} \hat n_s +U {\hat n}_1{\hat n}_{2} - \Delta (c_1^\dagger c_{2} + c_2^\dagger c_{1}).
\end{equation}
Here, $\hat n_s=c_s^\dagger c_{s}$ is the
electron number operator of quantum dot $s=1$ or $2$ with orbital energy $E_s$, 
and $U$ specifies the Coulomb interaction between two
dots.
The reservoirs of left and right ($\alpha=l$ and $r$)
electrodes are described by
\begin{equation}\label{eq2}
 H_R = \sum_{\alpha=l,r} H_{R_\alpha}
  = \sum_{\alpha=l,r}\sum_q \epsilon_{\alpha q} c_{\alpha q}^\dag c_{\alpha q} .
\end{equation}
The system--reservoirs coupling responsible for
electron transfer between the system and the electrodes is
\begin{equation} \label{eq3}
H_{SR}=\sum_{\alpha=l,r} \sum_{q}
 \left[T_{1q}^{(l)} c^\dag_1 c_{l q} + T_{2q}^{(r)} c^\dag_2 c_{r q} + {\rm H.c.} \right].
\end{equation}
The electron creation (annihilation) operators
$c^\dag_s$ $(c_s)$ and $c^\dag_{\alpha q}$ $(c_{\alpha q})$
involved in \Eqs{eq1}--(\ref{eq3})
satisfy the anti-commutator relations.
In this system, single electron transfer trajectories can be obtained
from the charge state of the DQD that is constantly measured by a
quantum point contact (QPC). Such a configuration was employed in
experiment \cite{Fuj061634} operated at a small bias voltage.

\subsection{Generalized non-Markovian rate equation}

We now turn to the non--Markovian rate equation.
Let $\rho(t)\equiv \mathrm{tr}_R\,\rho_T(t)$ be the reduced
system density operator.  The total density operator
is assumed to be initially factorisable into a system and a
reservoir part, $\rho_T(t_0)=\rho(t_0)\rho_R(t_0)$,
and the system--electrode couplings are assumed to be weak.
Using the standard approach,
one can readily derive a non-Markovian quantum master
equation.\cite{Wei99,Yan05187,Wel06044712}
For the present study, we adopt
$T^{(\alpha)}_{sq}=T^{(\alpha)}_s T^{(\alpha)}_q$ for simplification.
A rotating wave approximation to the cross
coupling terms between the system orbitals is not required here since each electrode
is coupled to one orbital site only.\cite{Wel08195315}
We denote the system Liouville operator $\mathcal L_{S}\cdot\equiv[H_{S},\cdot]$
and set $\hbar =1$.
The resulting quantum master equation
in the time-nonlocal form
reads \cite{Wel06044712,Wel08195315}
\begin{align}\label{equ:master2local}
\dot{\rho}(t)
= &  -i \mathcal L_S(t) \rho(t)
- \sum_{\alpha s} \int_0^t \mathrm d \tau \,\vert T^{(\alpha)}_s \vert^2 \nl\quad
& \times \Bigl\{C^{(+)}_{l}(t-\tau)
  \big[c_s, e^{-iH_S(t-\tau)} c_{s}^\dagger \rho(\tau) e^{iH_S(t-\tau)}\big]
 \nl
&\ \  - C^{(-)}_{l}(t-\tau) \big[
   c_s, e^{-iH_S(t-\tau)} \rho(\tau)  c_{s}^\dagger e^{iH_S(t-\tau)} \big]
\nl\quad &
+ \mathrm{H.c.} \Bigr\}.
\end{align}
The reservoir correlation functions,
\begin{equation}\label{equ:correl+}
C^{(+)}_{\alpha}(t)=\sum_{q} \vert T_{q}^{(\alpha)} \vert^2
  \langle c^{\dag}_{\alpha q}(t)c_{\alpha q}(0) \rangle_{R_\alpha},
\end{equation}
and
\begin{equation}\label{equ:correl-}
C^{(-)}_{\alpha}(t)=\sum_{q} \vert T_{q}^{(\alpha)}\vert^2
\langle c_{\alpha q}(0)c^{\dag}_{\alpha q}(t) \rangle_{R_\alpha},
\end{equation}
contain the properties of the electrodes.
Here, $c^{\dag}_{\alpha q}(t)\equiv
  e^{i H_{R_\alpha} t}  c_{\alpha q}^\dagger  e^{-i H_{R_\alpha} t}$
and $\la O \ra_{R_\alpha}\equiv \mathrm{tr}_{R_\alpha}
\lbrace O \rho_{R_\alpha} \rbrace$,
with $\rho_{R_\alpha}$ being the
density operator of the bare electrode $\alpha$
under a constant chemical potential $\mu_{\alpha}$.
Physically, $C^{(+)}_{\alpha}(t)$
describes the process of electron transfer
from the $\alpha$--electrode to the system,
while $C^{(-)}_{\alpha}(t)$ describes the reverse process.
These two correlation functions are not independent; they
are related via the fluctuation--dissipation
theorem.

Electron counting experiments are operated either
in the large--bias limit in order to achieve a {\em directional}
trajectory of single transfer events\cite{Lu03422,Fuj042343,Gus06076605}
or at small bias in order to realize transfer against
the direction of the bias.\cite{Fuj061634}
Non-Markovian effects are either due to small
bias or finite band-width.
In order to study both regimes, we derive a rate equation by
projecting the master equation (\ref{equ:master2local})
into the Fock space of system and by considering
only the population part $p_{m}\equiv \rho_{mm}$.
Some simple algebra leads from Eq.\,(\ref{equ:master2local}) to
the non-Markovian Pauli rate equation 
\begin{align}\label{ratecom}
\dot{p}_{m}(t) &= \sum_{\alpha n}\!
 \int_0^t \! \mathrm d \tau \big[
     \Gamma_{mn}^{(\alpha)} C^{(+)}_\alpha(t-\tau)e^{-i \omega_{mn} (t-\tau)}
     p_{n}(\tau)
\nl&\quad
 + \Gamma_{nm}^{(\alpha)} C^{(-)}_\alpha (t-\tau)e^{-i \omega_{nm} (t-\tau)}
       p_{n}(\tau)
\nl&\quad
  -\Gamma_{nm}^{(\alpha)} C^{(+)}_\alpha(t-\tau)e^{-i \omega_{nm} (t-\tau)}
    p_{m}(\tau)
\nl&\quad
  -\Gamma_{mn}^{(\alpha)} C^{(-)}_\alpha (t-\tau)e^{-i \omega_{mn} (t-\tau)}
    p_{m}(\tau)  \big] + {\rm c.c.}
\nl &\equiv
 \sum_n \int_0^t \! \mathrm d \tau K_{mn}(t-\tau) p_n(\tau).
\end{align}
Here, $\omega_{mn}\equiv E_m-E_n$ is
the transition frequency between two Fock states;
\be\label{eq:Gamn}
\Gamma_{mn}^{(\alpha)} = \vert T_s^{(\alpha)}
 \vert^2 \vert \bra{m} c^{\dagger}_s \ket{n}\vert^2,
\ee
with $s=1$ or $2$ for $\alpha=l$ or $r$, respectively,
is the state--dependent non-Markovian system--reservoir coupling strength.
As inferred from \Eq{eq:Gamn}, $\Gamma_{mn}^{(\alpha)}\neq 0$ only if
$\vert m\ra$ has one more electron than $\vert n\ra$.
We can therefore identify the {\it rate kernel}
elements involved in \Eq{ratecom} with
three physically distinct contributions
\begin{equation}\label{eq13}
K(t) \equiv \sum_{\alpha}[K^{(\alpha +)}(t)+ K^{(\alpha -)}(t)] + K_0(t).
\end{equation}
$K^{(\alpha +)}(t)$ and  $K^{(\alpha -)}(t)$ realize an electron
transfer in and out of the system through the
$\alpha$--electrode, respectively.
They summarize the off--diagonal matrix
elements of the transfer rate kernel $K(t)$
in \Eq{ratecom},
\begin{equation} \label{eq9}
K^{(\alpha +)}_{mn}(t)= \Gamma_{mn}^{(\alpha)}
  C^{(+)}_\alpha(t-\tau)e^{-i \omega_{mn} t} +{\rm c.c.},
\end{equation}
\begin{equation}\label{eq10}
K^{(\alpha -)}_{mn}(t)=\Gamma_{nm}^{(\alpha)}
 C^{(-)}_\alpha (t-\tau)e^{-i \omega_{nm} t} + {\rm c.c.}
\end{equation}
$K_0(t)$ summarizes the diagonal matrix elements of $K(t)$
and leaves the number of electrons in system unchanged.
These diagonal elements satisfy
\be\label{K0}
 (K_0)_{nn}=-\sum_{\alpha,m} \, \big[K^{(\alpha +)}_{mn}(t)+ K^{(\alpha -)}_{mn}(t)\big].
\ee

 For the Lorentzian spectral density model
[\Eq{Lorent_J}], where the reservoir
spectral density assumes the form 
$J_{\alpha}(\omega)=
\gamma^2_{\alpha}/[(\omega-\Omega_{\alpha})^2 +\gamma_{\alpha}^2]$,
we obtain for the off--diagonal rate kernel
elements the following expressions,
\begin{align} \label{KmnLoren}
K^{(\alpha \pm)}_{mn}(t)
&= 2 \Gamma_{mn}^{(\alpha)} \Big\{ e^{- \gamma_\alpha t}
 [a_\alpha^\pm \cos( \Omega^\alpha_{mn} t)
    - b_\alpha^\pm \sin( \Omega^\alpha_{mn} t)]
\nl &\quad
  + \sum_{k=1}^{\infty} e^{ -\varpi_k t}
 [c_{\alpha k}^\pm \cos( \mu^\alpha_{mn} t)
  - d_{\alpha k}^\pm  \sin( \mu^\alpha_{mn} t) ]\Big\}.
\end{align}
Here, $\varpi_k = (2k-1)\pi/\beta$
is the fermionic Matsubara frequency,
while $\Omega^\alpha_{mn}\equiv \Omega_\alpha-\omega_{mn}$ and
$\mu^\alpha_{mn}\equiv \mu_\alpha-\omega_{mn}$.
The coefficients $a_\alpha^\pm$,
$b_\alpha^\pm$, $c_\alpha^\pm$, and $d_\alpha^\pm$
are all real, given explicitly in Appendix\,\ref{Appendix1}
by \Eq{coef_all}.
The first term in the curly brackets of \Eq{KmnLoren} reflects
the spectral properties of the electrode-system coupling, while
the second term arises from the
decomposition into Matsubara frequencies which induces memory
effects due to small bias voltages. From the expressions one can infer
that large $\gamma_\alpha$, wide bands,
and large $\varpi_k$, high bias,
cause a fast decay of the transfer rates in time.
The decay is responsible
for the memory loss in the system.

The non-Markovian Pauli rate equation (\ref{ratecom}), in terms of
the population vector ${\bm p}(t)=\{p_m(t)\}$
and the involved transfer matrices, is 
\begin{equation}\label{rateeq}
\dot{\bm p}(t)=\int_{t_0}^{t} \mathrm d\tau \,
K(t-\tau) {\bm p}(\tau).
\end{equation}
It reads in Laplace frequency domain 
\begin{equation}\label{rateeq-laplace}
 s \tilde {\bm p}(s)- {\bm p}_0= \tilde K(s) \tilde{\bm p}(s).
\end{equation}
The corresponding electron transfer rates are
\begin{align}
\tilde K^{(\alpha \pm)}_{mn}(s) &
= 2 \Gamma_{mn}^{(\alpha)} \Bigl\{
  \frac{a_\alpha^\pm ( s+ \gamma_\alpha)- b_\alpha^\pm \Omega^\alpha_{mn}}
    {(s+ \gamma_\alpha)^2 +  (\Omega^\alpha_{mn})^2 }
\nl &\qquad\quad
 + \sum_{k=1}^{\infty}
  \frac{c_{\alpha k}^\pm (s+ \varpi_k)
    -d_{\alpha k}^\pm \mu^\alpha_{mn}}
  { (s+ \varpi_k)^2 +  (\mu^\alpha_{mn})^2 }
\Bigr\}.
\end{align}
The derived non-Markovian rate equation formalism
is based
on a microscopic description of the electrode-system coupling,
and is valid for arbitrary bias and temperature.
Compared to the quantum master equation
in the same regime \cite{Yan05187,Wel06044712},
the exclusion of the coherence makes it numerically
feasible to calculate multilevel
systems such as large molecules.\cite{Wel081137}
This allows to include non--Markovian effects
in large many--body systems,
e.g.\ quantum-chemistry calculations, since the properties
of the molecular-junction enter only through the couplings
$\Gamma_{nm}^{(\alpha)}$ and the fitting
parameters of the Lorentzian spectrum.

To rate equation (\ref{rateeq}), the Born-Markov approximation
can be applied by separating the integration variables and
extending the upper limit to infinity in \Eq{rateeq}.
The resulting integration over time,
\be
  W^{(\alpha \pm)}_{mn} = \int_{0}^{\infty}\!{\mathrm d}t\,
  K^{(\alpha \pm)}_{mn}(t)
   =\tilde K^{(\alpha \pm)}_{mn}(s) \vert_{s=0} ,
\ee
gives the Markovian electron transfer rates. The second identity
is via the Laplace domain rate equation (\ref{rateeq-laplace}),
by which the Born-Markov approximation amounts to the zero frequency
contribution.

\section{Non-Markovian waiting time distribution}
\label{secWTD}

\subsection{Statistics analysis}
We consider two consecutive electron transfers contained in a time
series as illustrated in \Fig{fig:0}. An electron entered the system
from the left electrode at an earlier time $t_0$ is detected at
time $t$ leaving the system through the right electrode. No other
electron transfers are detected in between.
The joint-probability for the consecutive electron transfer events is
\begin{equation}\label{joint1}
 P(t)=\lla  W^{(r-)} G(t,t_0) W^{(l+)} {\bm p}(t_0) \rra.
\end{equation}
Here, $\lla\cdots\rra$ denotes the sum over the final system states.
We assume that the transfer events are instantaneous compared
to the time-scale of the system propagation
in between as shown in \Fig{fig:0}.
Therefore we have used the Markovian forms of rate matrices,
for the ascribed two consecutive events.
This assumption is reasonable in accordance with electron
counting experiments,
where typical waiting times are long
compared to the fast transfer
events.\cite{Lu03422,Fuj042343,Gus06076605,Fuj061634}
The memory of the system is contained in
$G(t,t_0)$, the non-Markovian propagator of the system from $t_0$
to $t$ in absence of transfers.
It is therefore associated with the diagonal
rate matrix $K_0$ of \Eq{K0}, satisfying
\begin{equation}
\frac{d}{dt} G(t,t_0) = \int_{t_0}^{t} \mathrm d\tau'\,
 K_{0}(t-\tau') G(\tau',t_0) .
\end{equation}
For the given two--event case, the joint--probability
is equivalent to a waiting time distribution.\cite{Wel08195315,Wel081137}

Now consider the event of an electron transferred into the system and
the subsequent
waiting time before any other transfer takes place,
also referred to as survival probability.
In the present notation it is given by
$
\lla G(t,t_0) W^{(\alpha\pm)} {\bm p}(t_0) \rra.
$
While the joint probability is subject to the nature of the second transfer,
the specific form of the second event is
irrelevant to the survival probability.
Consequently, we introduce the survival time operator
\begin{equation}
Z^{(\alpha\pm)}(t,t_0)= G(t,t_0) W^{(\alpha\pm)}.
\end{equation}
If memory is absent, the
survival probability is indifferent from the previous
waiting times. To study the
memory of a previous survival time that carries on into the following
survival time, we introduce two--time joint survival probabilities
of the form
\begin{equation}\label{equ:twotimes}
 Q(\tau_2,\tau_1)=  \lla   Z^{(r-)}(\tau_2 , \tau_1)
   Z^{(l+)}( \tau_1, t_0) {\bm p}(t_0) \rra.
\end{equation}

\subsection{Non-Markovian corrections}

The formal solution to the propagator in Laplace domain is given by
\begin{equation} \label{prop1}
\tilde G(s)=\frac{1}{s-\tilde K_{0}(s)}.
\end{equation}
The complex Laplace frequency $s= \gamma + i \omega$ is
associated with the system residing in its state.
The bilateral Laplace transformation reduces to a Fourier transformation
by setting $\gamma=0$.
Since $\tilde K_{0}(s)$ is strictly diagonal in the many-body eigenspace
of the system, the matrix inversion required in \Eq{prop1}
can be efficiently carried out for large systems.

The technique of expanding the propagation into
non-Markovian corrections
has been applied to electron transport
recently.\cite{Bra06026805,Bra081745,Fli08150601}
Here we apply it to the WTD.
Let us first express \Eq{prop1} by its series
\begin{equation}
\tilde G(s) =\sum_{n=0}^\infty \frac{ [\tilde  K_{0}(s) ]^n} {s^{n+1}}.
\end{equation}
Assuming the derivative
$\partial^m_s [\tilde K_{0}(s) ]$ exists for all $m$,
the kernel can then be expanded into a Taylor series
$[\tilde K_{0}(s)]^n=\sum_{m=0}^\infty \partial^m_s
[\tilde K_{0}(s)]^n\vert_{s=0} \frac{s^m}{m!}$.
Thus,
\begin{equation}
\tilde G(s)= \sum_{n=0}^\infty \sum_{m=0}^\infty \frac{\partial^m_s
[\tilde K_{0}(s) ]^n\vert_{s=0}}{m!\, s^{n+1-m}}.
\end{equation}
Now we apply the inverse Laplace transform
$
x(t)=\frac{1}{2 \pi i} \int_{\gamma - i\infty}^{\gamma+i \infty}
\mathrm ds \, e^{st} \tilde x(s)
$
to switch back into time domain.
One can simplify the poles by using $m=n$
which neglects the transient terms.\cite{Bra06026805,Bra081745,Fli08150601}
We obtain
\be\label{eq18}
G(t) = \!\sum_{n=0}^{\infty}\! \frac{1}{n!}
\left[
  \frac{\partial^n}{\partial s^n}\!\!
  \left([ \tilde K_{0}(s) ]^n
    e^{\tilde K_{0}(s) t}
  \right)\right]_{s=0} \!\!
 \equiv \sum_{n=0}^{\infty} G^{(n)}(t),
\ee
with $G^{(n)}(t)$ denoting
the individual term involved,
where
$G^{(0)}(t)=e^{\tilde K_{0}(s) t} \vert_{s=0}$
describes the Markovian dynamics.
The first identity of expression (\ref{eq18})
is asymptotically exact since the dynamics
is reduced to the poles $m=n$.
The WTD
can also be expressed in terms
of $P(t)=\sum_n P^{(n)}(t)$,
with $P^{(0)}(t)$ denoting the Markovian contribution;
so can the survival probabilities.

\section{Demonstration and discussion}\label{demo}

We employ a non-Markovian rate equation to calculate the two--electron
system as illustrated in
\Fig{fig:0}. This system resembles the counting experiment
conducted in Ref.\,\onlinecite{Fuj061634}.
Here, the DQD provides a total number of four eigenstates:
the unoccupied ($\vert 0 \rangle$)
two single--occupied ($\vert 1 \rangle$ and $\vert 2 \rangle$),
and one double-occupied ($\vert 3 \rangle$),
with the energies of $\epsilon_0=0$,
$\epsilon_{1/2} =  \frac{1}{2}(E_1 + E_2)
 \mp \sqrt{\frac{1}{4}(E_1-E_2)^2 +  \Delta^2}
$, and $\epsilon_3=E_1+E_2+U$, respectively.
The equilibrium of the chemical potential of the electrodes
is set to $\mu^{\rm eq}=(E_1+E_2)/2$.
For numerical demonstrations, we use the numbers in
accordance with recent electron
counting experiments of electron transfers through
quantum dot systems at small temperatures.\cite{Gus06076605}
A coupling strength of $\Gamma= 10^4$\,Hz serves as the
unit for all values. This is equivalent to an
energy unit of $[E]=10^4 h= 6.63 \times 10^{-30}$\,J,
and a time unit of $[t]= 0.1$\,ms,
which is the typical time scale of waiting times in quantum dot counting
experiments.\cite{Fuj061634} We also use a low temperature of
$T=2 \times 10^{4} [E] = 10$\,mK.
If mentioned, we set a small energy
detuning of $\Delta E=E_1-E_2$ in order
to deduce specific frequencies of the systems.
The bandwidth $\gamma$ is set  sufficiently large in order
to neglect the finite bandwidth effects; thus
the non--Markovian effect is studied
in the wide band region.
In addition, the Lorentzian spectral densities are aligned
to the orbitals of the system.

\subsection{Transients and Fourier spectrum of WTD}

\begin{figure}
\includegraphics[width=8.5cm,clip]{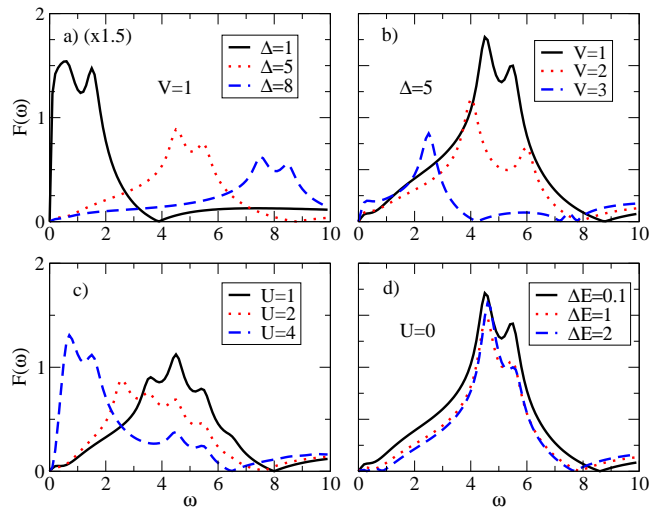}
\caption{The relative non-Markovian spectrum $F(\omega)$
of the WTD. The parameters used for
the four panels are given as follows.
Upper left panel (a): $U=0$, $\Delta E=0$, $V=1.0$, $\Delta = 1.0,5.0,8.0$.
Upper right panel (b):  $U=0$, $\Delta E=0$, , $\Delta = 5.0$. $V=1.0, 2.0, 5.0$.
Bottom left panel (c):  $\Delta E=0$, $V=1.0$, $\Delta = 5.0$ , $U=1.0, 2.0, 4.0$.
Bottom right panel (d):  $V=1.0$, $\Delta = 5.0$ , $U=0.0$,  $\Delta E=0.1, 1.0, 2.0$.
\label{fig:2}}
\end{figure}

Figure \ref{fig:2} shows the relative non-Markovian spectrum of the WTD represented by
\begin{equation}
  F(\omega)=\frac{1}{\Gamma}
 \frac{\vert P(\omega)-P^{(0)}(\omega)\vert }
   { P^{(0)}(\omega)}.
\end{equation}
It is noteworthy that $F(\omega)$ is independent of
the system reservoir coupling strength parameter $\Gamma$.
The WDT spectrum reveals several
frequencies that are present in
the transient oscillations.
These depend only on the
internal transfer rate $\Delta$,
Coulomb coupling $U$,
and bias voltage $V$.
The specific values of the parameters
are given in the caption of the figure.
Figure \ref{fig:2}(a) shows the main characteristics
of $F(\omega)$, consisting of
two overlapping sub-peaks centered around the value of $\Delta$.
Changing the value of $\Delta$
leads to the shift of both sub-peaks equally by
$\Delta$. In \Fig{fig:2}(b), $\Delta$ is kept constant
and the bias voltage is varied. We find that
the splitting of the two sub-peaks
is determined by the applied voltage.
Labeling the two peaks with $\pm$, respectively, we can
deduce the following relation for the
corresponding characteristic frequencies.
$\omega_{\pm}= \vert \Delta \pm V/2 \vert$.
In the presence of Coulomb interaction,
we observe an additional double--peaks
feature at $\vert U -\Delta \pm V/2 \vert$,
as demonstrated in \Fig{fig:2}(c).
This is similar to a non-Markovian shot noise
spectrum,\cite{Jin0806}
where a
finite Coulomb interaction $U$ induces also additional peaks
due to the
energy gap between the two-particle occupation state
and lower states.
On the other hand,
the orbital detuning does not induce additional
peaks in the double quantum dot in series
as shown in \Fig{fig:2}(d).

 Oscillations of Rabi frequency, which
were observed in parallel DQD
systems,\cite{Wel08195315,Wel0957008,Bra08477}
are however absent in the present series DQD system.
In the parallel cases, the transport proceeds
via two channels, and the Rabi oscillations
in the WDT can be observed
as the consequence of quantum
mechanical interferences.\cite{Wel08195315,Wel0957008,Bra08477}
It is also noted that the information
contained in the spectrum of the non-Markovian
WDT is mostly equivalent
to a measurement of the
non-Markovian shot noise spectrum.
For this purpose, the WTD can be
considered as an alternative approach
to the shot noise spectrum measurement.

\subsection{Memory landscape of consecutive waiting times}

\begin{figure}
\begin{center}
\includegraphics[width=8.0cm,clip]{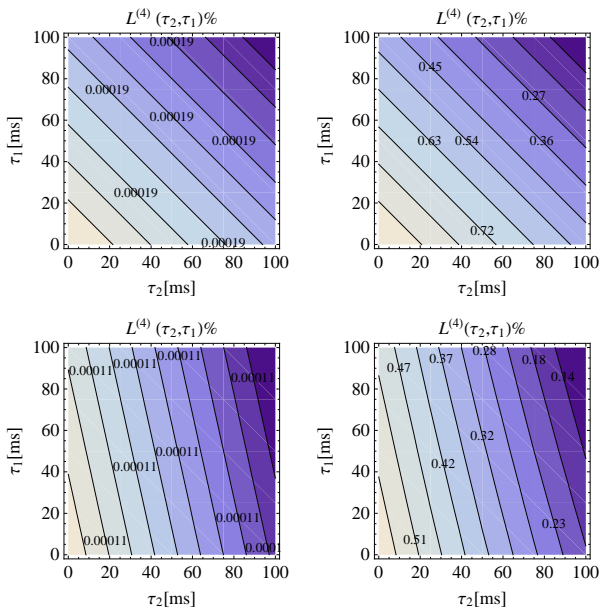}
\end{center}
\caption{$L^{(4)}_{l,r}$ memory landscape of
consecutive survival times.
The bandwidth is large and a finite bias of $V=0.1 k_b T$ is applied.
The left coupling strength is $\Gamma^{(l)}=10^{4} Hz$.
The upper panels are calculated in absence of Coulomb coupling, $U=0$, bottom
panels display the Coulomb blockade regime, $U= \infty$. 
Left panels are calculated for
a symmetric system, $\Gamma^{(l)}=\Gamma^{(r)}$. In the right panels, 
a stronger coupling strength is applied to the right electrodes
$\Gamma^{(r)} = 10^{6} Hz$. 
\label{fig:3}}
\end{figure}

The expansion in non-Markovian corrections, \Eq{eq18},
 can be readily employed to
calculate the two propagators involved in
the two-times joint probabilities
defined by \Eq{equ:twotimes}.
Denote
\begin{align}
 Q^{(n)} (\tau_2,\tau_1) &=
 \sum_{k=0}^{n} \sum_{j=0}^{n} \lla
 G^{(k)}(\tau_2-\tau_1) W^{(r-)}
\nl &\qquad\ \ \times
 G^{(j)}(\tau_1-t_0) W^{(l+)}
 \bm p(t_0) \rra.
\end{align}
A memory landscape of the system
can be calculated by the difference
between non-Markovian and Markovian
two-times joint probabilities
\begin{equation}\label{mem-land}
L^{(n)} (\tau_2,\tau_1)=
 \frac{ Q^{(n)} (\tau_2,\tau_1)
 - Q^{(0)} (\tau_2,\tau_1) }
  {Q^{(0)} (\tau_2,\tau_1)}.
\end{equation}
The order $n$ of the perturbative expansion in non-Markovian
corrections has to be chosen in accordance to the parameters
in order to assure satisfactory convergence.
We find that the summation to the fourth non-Markovian contribution already converges
sufficiently for the given parameters.
As the memory in \Eq{mem-land} decays,
the relative non-Markovian landscape
$L^{(n)} (\tau_2,\tau_1)$ converges to zero.
Figure\,\ref{fig:3} shows the memory landscape of two survival times
related to two consecutive electron transfers through the left
electrode.
It visualizes how memory of the waiting time $\tau_1$
after the first transfer
is carried over into the waiting time $\tau_2$
following the second transfer.

We find that the non-Markovian effects
are small for the given parameters in case the DQD is coupled symmetrically
to the electrodes. This is due to the relatively weak coupling of the DQD to the electrodes
which is required in present counting experiment in order to
resolve single electron transfers on the measurable timescales.
It is observed that a stronger coupling to only one electrode
induces significantly larger non-Markovian
effects as shown in the left panels of \Fig{fig:3}.

In general, fast electron transfers
are necessary in order to observe significant non-Markovian effects.
The deviations for $\tau_1$, $\tau_2$ approaching zero from the Markovian
value are due to the truncation of the transients in the derivation
of the expansion. The expansion follows the general trend
of a numerically exact solution. Both solutions
overlap after the transients have decayed. However this causes
relatively large inaccuracies for $\tau_1$ and $\tau_2$ close to zero.

There is an interesting dependency of the non-Markovian effects in
the memory landscape on the Coulomb repulsion $U$.
By comparing the upper panels of \Fig{fig:3} where Coulomb repulsion is absent,
with the bottom one, where a large $U$ induces a Coulomb blockade regime,
we observe that memory decays faster with $\tau_2$
in the Coulomb blockade regime. This can be explained as follows.
In the second regime, only a single electron can
occupy the DQD and the double occupancy state does not provide memory
for the second survival time leading to an overall smaller non-Markovian
contribution. In this case, only one possible trajectory in the left to right direction is
possible. An electron enters the unoccupied DQD at time $\tau_1=0$ and leaves it at time $\tau_2=0$.

The memory is preserved during $\tau_1$ by the single electron inside the DQD.
However, after the electron has left a junction, the memory of its trajectory is lost rapidly
since no other electron can serve as a messenger inside the DQD
thus leading to comparatively short survival times $\tau_2$ where memory is present.
In the regime where Coulomb repulsion is neglectible, a second electron can occupy the
junction along the described trajectory, which is represented in the model by the presence
of an occupied double occupancy state. The presence of the second electron preserves
the memory during $\tau_2$ after the other electron has left the junction.

\section{Conclusion}

We find that non-Markovian effects are small in the regimes
of recent single electron counting experiments.
The sampling rate of current experiments is slow, a requirement which is
imposed by the detection process of single electron transfers with currently available
technology. This verifies the reason that the Markovian approximation of
previous studies considering FCS or WTD
is reasonable for the previously investigated systems.

Non-Markovian effects in the electron transfer statistics
have to be taken into consideration for stronger
electrode-DQD couplings, which then also requires faster sampling
rates or a strongly asymmetric system.
For example they affect the decay rates of the WTD
which are directly related to the electronic
structure of the system in junction.\cite{Wel081137}
Non-Markovian effects also induce several oscillations with
characteristic system frequencies.

{Note that the form of Pauli rate equation remains valid 
itself in the strong coupling limit. In the present paper 
we employ a perturbative approach to the rate 
equation and observe that the non-Markovian
effects increase with the coupling strength. This observation
is expected to remain true based on general Pauli rate equation
dynamics. In other words, the non-Markovian effects
are mainly visible for stronger couplings.} 

The employed microscopic non-Markovian rate equation provides
a general framework to study similar systems and allows us to distinguish
between non-Markovian effects due to intrinsic properties of the
system, finite electrode-system coupling band-width and small bias voltages.
It can be combined with quantum chemistry calculations that can
calculate the employed parameters for molecules and their binding to
the electronic bands of the metal electrodes.
The approaches derived for the non-Markovian WTD are general
and can be applied to a variety of processes in physics,
chemistry and biology that are described by rate equations.

\acknowledgments
 Support from the RGC (604007 \& 604508) of Hong Kong
is acknowledged.

\appendix*

\section{Rate coefficients}
\label{Appendix1}

Introducing the coupling reservoir spectrum density
$J_{\alpha}(\omega)=\pi \sum_q \vert T^{(\alpha)}_q \vert^2
\delta(\omega-\epsilon_{\alpha q})$
and applying Fermi statistics to the reservoir modes,
the correlation functions (\ref{equ:correl+}) and (\ref{equ:correl-})
can be written as
\begin{equation}\label{FDT}
 C_{\alpha}^{(\pm)}(t)=\int_{-\infty}^\infty \frac{\mathrm d\omega}{\pi} \,
 J_{\alpha} (\omega) \,f_\alpha^{(\pm)}(\omega)  e^{\mp i \omega t}.
\end{equation}
Here,
$f_\alpha^{(+)}(\omega)=1-f_\alpha^{(-)}(\omega)
=[1+e^{\beta(\omega-\mu_{\alpha})}]^{-1}$ is
the Fermi distribution function, with $\beta=1/k_b T$ being
the inverse temperature and $\mu_{\alpha}$
the chemical potential of to the $\alpha$--electrode.
Adopting a Lorentzian form of spectral density,
\begin{equation}\label{Lorent_J}
 J_{\alpha}(\omega)=
\gamma^2_{\alpha}/[(\omega-\Omega_{\alpha})^2 +\gamma_{\alpha}^2],
\end{equation}
the finite spectral width parameter
$\gamma_{\alpha}$ is used to characterize the
non-Markovian nature of system--reservoir
coupling. With the complex roots of the Fermi
function and of the Lorentzian
spectral density, the integrals in \Eq{FDT} can be determined
by the residues of the Kernel. The resulting infinite series are
\be\label{appC+}
 C_{\alpha}^{(+)}(t)=  \gamma_\alpha
  f_{\alpha}^{(+)}(\Lambda_{\alpha}) e^{ i\Lambda t}
-\frac{2i}{\beta} \sum_{k=1}^{m}
 J_{\alpha}(\upsilon_k) e^{ i \upsilon_k t}
\ee
and
\be\label{appC-}
C_{\alpha}^{(-)}(t)
= \gamma_\alpha f_{\alpha}^{(-)}(-\Lambda_{\alpha}^\ast)
  e^{- i\Lambda^\ast t}
-\frac{2i}{\beta} \sum_{k=1}^{m}
 J_{\alpha}(\upsilon_k^\ast) e^{- i \upsilon_k^\ast t},
\ee
with the abbreviation $\Lambda_\alpha=\Omega_\alpha+i \gamma_\alpha$
and $\upsilon_k=\mu_\alpha
 +i\varpi_k$, where
$\varpi_k\equiv
 (2k-1)\pi/\beta$ are
the Fermion Matsubara frequencies.
In order to completely separate real and imaginary parts of the correlation functions
(\ref{equ:correl+}) and (\ref{equ:correl-}), we first separate its individual
components.
For the two complex Fermi functions we calculate
\begin{align}
  f_{\alpha}^{(\pm)}(\pm\Omega_{\alpha}+i \gamma_{\alpha})
= \frac{1+e^{\pm\beta(\Omega_{\alpha} -\mu_{\alpha})}
 e^{-i\beta\gamma_{\alpha}}}{X^{\pm}_{\alpha}},
\end{align}
where
\be
  X^{\pm}_{\alpha} \equiv 1+ 2\cos (\beta\gamma_{\alpha})
    e^{\pm \beta( \mu_{\alpha}-\Omega_{\alpha})}
   + e^{\pm 2\beta( \mu_{\alpha}-\Omega_{\alpha})}.
\ee
The complex spectral densities can be separated into
\begin{align}
&\quad J_{\alpha}(\mu_{\alpha}\pm i \varpi_k )
\nl &=
 \frac{[\varpi_k^2 +(\mu_{\alpha}-\Omega_{\alpha})^2 +\gamma_{\alpha}^2]
   \mp 2 i \varpi_k (\mu_{\alpha}-\Omega_{\alpha})}
  {Y_{\alpha}},
\end{align}
where
\be
  Y_{\alpha} \equiv \frac{1}{\gamma^2_{\alpha}}
  [\varpi_k^2 +(\mu_{\alpha}-\Omega_{\alpha})^2 +\gamma_{\alpha}^2]^2
   + 4 \varpi_k^2 (\mu_{\alpha}-\Omega_{\alpha})^2.
\ee
Based on the separation in real and imaginary contributions,
we can write the correlation functions as
\begin{align} \label{decomp3}
 C_{\alpha}^{(\pm)}(t)
&= (a_{\alpha}^\pm +i b_{\alpha}^\pm)
  e^{ (\pm i\Omega_{\alpha}-\gamma_{\alpha}) t}
\nl &\quad
  +\sum_{k=1}^{m} (c_{\alpha k}^\pm + i d_{\alpha k}^\pm)
   e^{ (\pm i\mu_{\alpha}-\varpi_k) t} .
\end{align}
The coefficients are all real:
\bsube\label{coef_all}
\begin{align}
a_{\alpha}^\pm  &=  \frac{\gamma_{\alpha}}{X^{\pm}_{\alpha}}
  [1+ e^{\pm \beta( \Omega_{\alpha} -\mu_{\alpha})}
  \cos (\beta \gamma_{\alpha})] ,
\\
b_{\alpha}^\pm &= \frac{\gamma_{\alpha}}{X^{\pm}_{\alpha}}
  [e^{\pm \beta( \Omega_{\alpha} -\mu_{\alpha})}
  \sin (\beta \gamma_{\alpha})],
\\
c_{\alpha k}^\pm &= \mp
 \frac{4(\mu_{\alpha}-\Omega_{\alpha})\varpi_k  }{\beta Y_{\alpha}},
\\
d_{\alpha k}^\pm &= \mp
 \frac{2[\varpi_k^2 +(\mu_{\alpha}-\Omega_{\alpha})^2 +\gamma_{\alpha}^2] }{\beta Y_{\alpha}}.
\end{align}
\esube
Rigorously, the sum over the Matsubara values would be infinite;
i.e., $k_{\rm max} = m\rightarrow \infty$ in \Eqs{appC+} and (\ref{appC-}),
but it can be truncated for practical purposes at a finite value
that depends on the temperature of the system $T$ and the spectral width.



\end{document}